# A Formal Definition of Model Composition Traceability

Youness Laghouaouta[1], Adil Anwar[2] and Mahmoud Nassar[1]

[1] IMS-SIME, ENSIAS, Mohammed V University of Rabat
Rabat, Morocco

[2] Siweb, EMI, Mohammed V University of Rabat
Rabat, Morocco

**Abstract**

In a multi-modeling based approach, the system under development is described by several models that represent various perspectives and concerns. Obviously, these partial representations are less complex than the global model, but they need to be composed to address validation and synchronization tasks. The model composition is a crucial model driven development operation, but it remains a tedious and error prone activity. In this perspective, a traceability mechanism offers a way to master this complexity by providing support to comprehend the composition effects. In previous work, we presented a traceability approach dedicated to this operation. The current takes advantages of these experiments, and proposes a formalization of the model composition traceability. Also, an overview of a generic traceability approach is provided. The latter relies on the formal definition we introduce for the model composition operation and the related traces.

## 1. Introduction

The Model Driven Engineering (MDE) advocates the use of models to represent any artifacts handled by the software development process (e.g. requirements, design products, source code, etc.). The principle is to provide an abstract representation of these artifacts, by means of the meta-modeling technique, that can be manipulated by machines. Thus, the final system is produced by a succession of model management operations that are applied on these models. In addition to that, to reduce the complexity of this activity, developers can describe the system by as many models as they want. These partial representations are less complex than the global model. Each one refers to a specific concern, represents a particular perspective, or describes a given component.

Nevertheless, multi-modeling approaches require a composition support. The model composition operation provides a mechanism for the validation of partial models, their synchronization, and the understanding of existing interrelations. But, despite of its benefits that simplify a software development both modular and driven by models, the composition operation remains a laborious and error prone activity.

In this perspective, a traceability mechanism offers a way to master this complexity. Indeed, traceability information exposes the exact effects of executing the composition operation, and helps understanding the relationships existing among partial models. However, through the examination of model traceability approaches proposed in the literature, we did not encounter a specific solution addressing the composition of models. Therefore, adopting an existing approach compromises expressiveness and reusability of the obtained traces.

To address this problem, we presented in previous work a traceability approach dedicated to the model composition operation [1][2]. Two languages have been targeted: the dedicated merging language EML [3], and the transformation language ATL [4]. The current work takes advantages of these experiments and proposes a formalization of the model composition traceability. It introduces a formalism that represents the model composition operation, as well as a formal definition of the corresponding traces and the generation mechanisms. Besides, this formalization effort constitutes a basis for the definition of a generic traceability approach independent form a given composition language.

The remainder of the paper is structured as follows. Section 2 presents a case study that will illustrate concepts of the proposed formalization. The latter will be detailed in Section 3 by introducing a formal definition of the model composition operation and the related traceability information to be captured. Section 4 gives an overview of our generic traceability approach. Related works are discussed in Section 5. Finally, a conclusion is given in Section 6.

## 2. Illustrative Example

This section presents the composition scenario we use to illustrate the core concepts of the proposed formalization. We will start by introducing the involved models. Thereafter, we will present the EML and ATL specifications that allow producing the expected composed model. Finally, the corresponding trace model will be given for a better understanding of the traces structuring.

2.1 Composition scenario

The composition scenario we have chosen is the merging of a system of entities with a vocabulary of terms. This example can be downloaded from the Epsilon platform web site[1]. Fig. 1 depicts the two source models and the model resulting from their composition. Each entity of the left model is merged with its corresponding term in the opposite model (the entity's name must match the name of the term or one of its aliases). It follows from their merging the production of a target entity having the name of its equivalent term. As for other entities that have no corresponding, they are simply copied to the composed model.

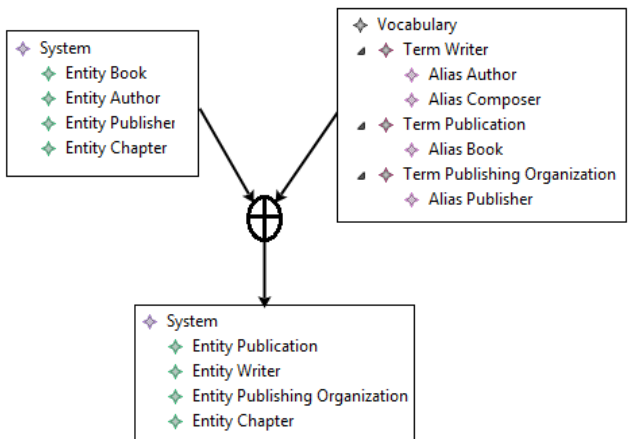

Fig. 1 The selected composition scenario.

2.2 Composition specification

In this section, we present the EML and ATL specifications that allow the realization of the selected merging scenario.

*a) Specifying the composition in EML*

The composition operation is specified in the Epsilon platform [5] by two separate scripts. The first comprises comparison rules (written in ECL [5]), and makes it possible to establish correspondences between the source models to be composed. The second script is an EML [3] specification which expresses the composition behavior. Listing 1 presents the comparison module. Its first rule (*MatchSystemWithVocabulary*) connects each system of the left model to an entity of the right model. These elements are captured by the **match** and **with** clauses. The second rule (*MatchEntityWithTerm*) compares entities with terms. The comparison logic is specified in the **compare** clause (Listing 1, lines 11-13) which checks if the name of the entity corresponds to the term's name or is among its aliases.

---
[1] http://www.eclipse.org/epsilon/examples/

Listing.1 ECL module implementing the comparison step

1.  rule MatchSystemWithVocabulary
2.  match s : Source!System
3.  with v : Vocabulary!Vocabulary {
4.  compare {
5.         return *true*;}
6.  }
7.
8.  rule MatchEntityWithTerm
9.  match s : Source!Entity
10. with t : Vocabulary!Term {
11. compare {
12. return s.name = t.name or t.`alias`.exists(a|a.name = s.name);}
13. }

Listing 2 illustrates a merging module. This specification includes two types of rules: merge rules which are identified by the syntax "**rule** *ruleName* **merge** *LeftParam* **with** *RightParam* **into** *TargetParam*", and transformation rules ("**rule** *ruleName* **transform** *SourceParam* **to** *TargetParam*"). Both *MergeEntityWithTerm* and *MergeSystemWithVocabulary* rules encapsulate a merging behavior, and they are applied on pairs of corresponding elements. They respectively merge entities with the corresponding terms and combine systems with vocabularies.

Listing.2 EML specification to realize the composition scenario

1.  rule MergeEntityWithTerm
2.  merge s : Source!Entity
3.  with t : Vocabulary!Term
4.  into m : Target!Entity {
5.  m.name = t.name;
6.  m.inDomain = *true*;}
7.
8.  rule MergeSystemWithVocabulary
9.  merge s : Source!System
10. with v : Vocabulary!Vocabulary
11. into t : Target!System {
12. t.entity = s.entity.equivalent();}
13.
14. rule TransformEntity
15. transform s : Source!Entity
16. to t : Target!Entity {
17. t.name = s.name;
18. t.inDomain = *false*;}

Elements to be merged are captured by the parameters **merge** and **with**, whereas the clause **into** specifies the produced element. An activation of the rule *MergeEntityWithTerm* constructs a merged entity. While the second merging rule (*MergeSystemWithVocabulary*) produces the composed system, and connects it to the target entities (Listing 2, line 12). The access to output elements of a rule activation can be performed through the *equivalent*() operation [6]. Indeed, the latter allows the resolution of target equivalents corresponding to a source model element. Finally, the rule *TransformEntity* is applied to source entities do not match any term of the vocabulary. It transcribes these entities into the composed model.

### b) Specifying the composition in ATL

The ATL language being dedicated to the model transformation operation makes more complex and less intuitive the specification of composition scenarios. Listing 3 presents the ATL module implementing the presented merging scenario. We notice the integration of all specific tasks (comparison and composition) in the same specification.

Listing. 3 ATL module implementing our composition scenario

```
1.  module merge;
2.  create OUT : Source from IN : Source, IN1 : Vocabulary;
3.  helper context Source!System def:
    match(r:Vocabulary!Vocabulary): Boolean=true;
4.  helper context Source!Entity def: match(r:
    Vocabulary!Term) : Boolean = self.name = r.name or
    r.alias->exists(a|a.name = self.name);
5.  --isLeft() : verifies if the source element belongs to the left
    model
6.  --isRight(): verifies if the source element belongs to the
    right model
7.  --getRightCorresp(): return the corresponding of a left
    element.
8.
9.  rule MergeEntityWithTerm{
10.    from s:Source!Entity, t:Vocabulary!Term (s.match(t)and
       s.isLeft()and t.isRight())
11.    to m : Source!Entity
12.    do {
13.    m.name <- t.name;
14.    m.inDomain <- true;}}
15.
16. rule MergeSystemWithVocabulary{
17.    from s:Source!System, v:Vocabulary!Vocabulary
       (s.match(v)and s.isLeft()and v.isRight())
18.    to t : Source!System
19.    do {
20.    for(x in s.entity->asSequence()){
21.    if(x.getRightCorresp().oclIsUndefined()){
22.          t.entity<-thisModule.resolveTemp(x,'m');}
23.    else{
24.    t.entity<-thisModule.resolveTemp(Tuple
       {s=x,t=x.getRightCorresp()},'m');}
25.       }
26. }}
27.
28. rule TransformEntity{
29.    from s : Source!Entity (s.isLeft() and
       s.getRightCorresp().oclIsUndefined())
30.    to m : Source!Entity
31.    do {
32.    m.name <- s.name;
33.    m.inDomain <- false;}}
```

The two helpers *match*() (Listing 3, lines 3-4) express the logics for comparing systems with vocabularies and entities with terms. These helpers are used to restrict the application of the rules *MergeEntityWithTerm* and *MergeSystemWithVocabulary* (Listing 3, lines 9-14, 16-26) on elements having a corresponding in the right model. On the other hand, the resolution of the container model (right or left model) is carried out by means of the *isLeft*() and *isRight*() helpers.

Unlike EML, ATL does not allow the categorization of rules on merging and transformation rules. For example, the rules *MergeEntityWithTerm* (Listing 3, lines 9-14) and *TransformEntity* (Listing 3, lines 28-33) implement two different behaviors, but no distinction is made regarding their expression. In both cases, target elements (determined by the clause **to**) are produced by transforming source elements captured by the parameters belonging to the clause **from**. Since these parameters select all elements that correspond to their types, the helpers operations *match*(), *isLeft*(), *isRight*(), and *getRightCorrespond*() are exploited to switch rules applications according to specificities of the composition operation (Listing 3, lines 10,17,29). As for a rule body, it expresses the way in which target elements are constructed. In general case, its specifies initialization for properties of target elements, and describes the weaving of structural relationships among composed model elements (mainly by using the *resolveTemp*() operation which resolves the target equivalent of a given source element) (Listing 3, lines 22,24).

## 2.3 The corresponding trace model

Fig. 2 depicts the trace model corresponding to the composition scenario we introduced in Section 2.1 (taking account of the EML specification). The generation mechanisms will be gradually presented in the following section, and we claim by presenting this model to illustrate traces we seek to capture. It should be noted that we used the Emf2gv project [1] to build a human friendly representation for traces. The dashed blue, green, and red lines reference, respectively, elements belonging to the left, right, and composed models.

The generated trace model comprises two types of traceability links: merging links (*MergingLink* nodes) and transformation links (*TransformationLink* nodes). These links capture the different composition correspondences. For example, the root merging link connects the left and right models (corresponding to the system and the vocabulary of terms) with the model resulting from the composition (the composed system). While the contained transformation link keeps track of the transcript of the

---
[1] http://emftools.tuxfamily.org/wiki/doku.php?id=emf2gv:start

*Chapter* entity to the composed model. As for its other descendants, they represent the merging of the *Author*, *Publisher* and *Book* entities with their equivalent terms.

The relations drawn up by solid lines represent the nesting of traceability links. This information specifies, for example, that the merging of the *System* and the *Vocabulary* elements initiated the transformation of the *Chapter* entity.

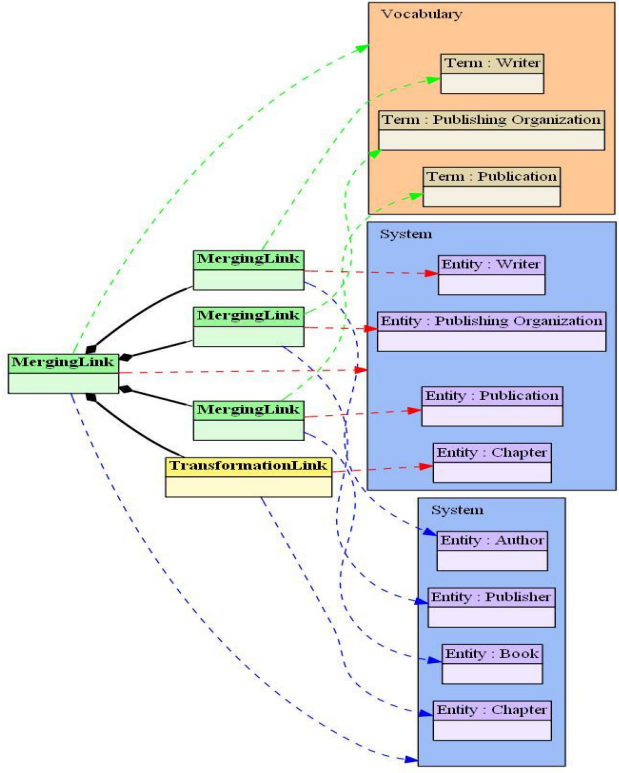

Fig. 2 The generated trace model.

## 3. Formalization of the model composition operation and the corresponding traces

This section starts by rigorously defining the elements constituting a composition specification, and will be followed by the description of traceability information to be captured. We note that this proposal does not provide a complete formalization of the composition operation and its traces. Our objective is to bring a better visibility on our traceability approach without producing a complete formalization that takes account of all its aspects.

3.1 Formalization of the composition operation

In the literature, model composition is generally defined as an operator that is applied to a set of source models to produce the corresponding composed ones, without any restriction on the number of the implied models. However, the composition operator is practically considered as a binary operator that combines the contents of two source models (the left and right models) into a single composed model. In our context, the composition generates in addition to the default target elements, other elements expressing the related traces. Therefore, we are expanding the number of target models in a maximum of two (the composed model and the trace model).

*Definition 1: Composition operator*
Let $Comp$ be a composition operator. The latter is applied on a left model $M_{left}$ (conforms to a metamodel $MM_{left}$) and a right model $M_{right}$ (conforms to a metamodel $MM_{right}$) to generate composed models $M_{comp_i}$ (each one conforms to a metamodel $MM_{comp_i}$).

$$\{M_{comp_i}\} = Comp(M_{left}, M_{right}) \quad i \in \{1,2\} \quad (1)$$

A composition specification expresses the way in which the composition operator is applied on the source models. It takes the form of a set of composition rules (essentially in rule-based approaches). Each rule defines a selection and, eventually, filtration mechanisms for elements on which the rule is applied, as well as actions to perform the composition. The selection activity is based on types of the involved model elements (specified in the input parameters of the rule). While the filtration mechanism can be explicitly specified by a set of conditions (e.g. a filter to associate with an input pattern element of an ATL rule), or based on predefined composition strategies exploiting the comparison relationships (e.g. to restrict the application of EML merging rules on elements having a corresponding in the opposite model). On the other side, and since the composition process implements three basic operations (comparison, merging, and transformation), the body of each rule expresses elementary actions which are intended to allow one of these behavior.

*Definition 2: composition specification*
A composition specification $Spec$ consists of a set of composition rules $R_{comp}$:

$$Spec \stackrel{\text{def}}{=} \{R_{comp_i} | i \in \mathbb{N}^*\} \quad (2)$$

*Definition 3: Composition rule*
Let $R_{comp}$ be a composition rule. It is defined by the following quintuplet:

$$R_{comp} \stackrel{\text{def}}{=} (nom, T_{in_{left}}, T_{in_{right}}, T_{out}, type) \\ type \in \{\text{"match"}, \text{"merge"}, \text{"transform"}\} \quad (3)$$

- $nom$ : *The rule name.*
- $T_{in_{left}}$ : *Types of the left input parameters. These types are defined in the left metamodel $MM_{left}$.*
- $T_{in_{right}}$ : *Types of the right input parameters. These types are defined in the right metamodel $MM_{right}$.*
- $T_{out}$ : *Types of the target parameters. These parameters reference the elements to produce, and the corresponding types are defined in the metamodels $\{MM_{comp_i} \mid i \in \{1,2\}\}$.*
- $type$ : *The rule type.* "match" *for a comparison rule,* "merge" *for a merging rule, and* "transform" *to mark transformation rules.*

During the execution of a composition specification, its different rules are applied on the source model elements, which are captured by the input parameters, in order to produce the composed model elements. Rules applications are conducted according to a predetermined order, which may be: (i) explicitly specified by the developer (e.g. express a rule invocation); (ii) derived from the structure of the source models (e.g. start by activating rules applicable on the first elements founded in a XML tree); (iii) driven by rules priorities (e.g. start the execution of an ATL module by the *entrypoint* rule [4]); (iv) based on pre-defined strategies (e.g. apply first merging rules in EML). Hence, the composed models are generated iteratively, and their elements are produced in parallel to the rules applications. We call the application of a rule on certain source model elements "*rule activation*". We note by $Act_R$ the set of activations of a composition rule $R$ that have been launched for the realization of a given composition scenario.

***Definition 4: Execution of a composition specification***
We note by $Act_{Spec}$ all activations of composition rules which define a composition specification $Spec$ :

$$Act_{spec} \stackrel{\text{def}}{=} \{ a \mid a \in Act_R \land R \in Spec \} \quad (4)$$

***Definition 5: Composition rule activation***
Let $a$ be an activation of a composition rule $R$. This activation consists of a correspondence between the source model elements on which the rule is applied (including right model elements $El_{right}$ and left model elements $El_{left}$), with the target elements that have been added to the composed model $El_{comp}$.

$$a \in Act_R \stackrel{\text{def}}{=} (R, El_{left}, El_{right}, El_{comp}) \quad R \in Spec \quad (5)$$

***Definition 6: Sub-set of model elements***
We note by $El_x$ a sub-set of element belonging to a model $M_x$. For each element of the set $El_x$, we have:

$$el_x \in El_x \Rightarrow el_x \in M_x \quad (6)$$

One of the actions that can be expressed in the body of a composition rule is the call to another rule (e.g. a call to an imperative rule specified in an ATL module). The main objective through this mechanism is to weave structural relationships between the elements produced by the activation of the calling rule and those generated by the called rule. An explicit call is defined by the name of the called rule, and it specifies the source elements on which the rule should be applied. In what follows, we note an explicit call to a composition rule $R$ by $appelExp_R$.

***Definition 7: Explicit rule call***
Let $appelExp_R$ be an explicit call to a composition rule $R$. This call requests the application of the rule $R$ on a set of source elements $El_{left}$ and $El_{right}$ during a given activation of the calling rule $R_0$.

$$appelExp_R \stackrel{\text{def}}{=} (activ, nom, El_{left}, El_{right}) \quad where$$
$$\exists R_0, R \in Spec, activ \in Act_{R_0} \land R.nom = nom \quad (7)$$

Another rule call mechanism can be expressed, and it allows a current activation to use correspondences established by previous rule activations. Indeed, during the activation of a rule $R_0$, the resolution of target equivalents corresponding to a given source model elements (denoted by $resolvedEl$) can be requested. This is performed by means of a specific operation, we name $resolve$ (e.g. the *equivalent*() operation in EML, the *resolveTemp*() operation provided by ATL). We note an implicit rule call by $appelImp$.

***Definition 8: Implicit rule call***
An implicit call to a composition rule ($appelImp$) is defined by:

$$appelImp \stackrel{\text{def}}{=} (activ, resolve, resolvedEl) \quad where$$
$$(\exists R_0 \in Spec, activ \in Act_{R_0}) \land resolvedEl \subset M_{left} \cup M_{right} \quad (8)$$

***Definition 9: Target equivalents resolution***
The $resolve$ operation is a mapping between the source elements to resolve and their equivalents in the composed model

$$resolve : (M_{left} \cup M_{right})^* \to (\cup M_{comp_i})^*$$
$$source \to target \quad (9)$$
$$target = \{a.El_{comp} \mid a \in Act_{spec} \land source = a.El_{left} \cup a.El_{right}\}$$

## 3.2 Formalization of model composition traces

Traceability links that capture the effects of executing a composition specification are recorded in a separate model. The latter is generated as a product of the composition. In addition, two categories of links have to be expressed according to the composition rules types (comparison rules will not be traced): merging links to keep track of merging rules activations, and transformation links for transformation rules activations. Moreover, these traceability links should be connected to each other in order to express the rule calls sequence (explicit and implicit calls).

*a) Traces structuring*

**Definition 10: Trace model**
Let $M_{trace}$ be a trace model. This model is defined by the following pair:

$$M_{trace} \stackrel{\text{def}}{=} (L_{tr}, R_{tr}) \quad (10)$$

- $L_{tr}$ : Represents the set of traceability links. We set $L_{tr} = L_{merge} \cup L_{transf}$, where $L_{merge}$ is the set of merging links and $L_{transf}$ is the set of transformation links.
- $R_{tr}$ : Represents the set of relationships connecting traceability links.

A traceability link captures the effects of a given composition rule activation. This activation produces a set of composed model elements from certain right and left elements.

**Definition 11: Traceability link**
We define a traceability link $(l_{tr})$ by a triplet that keeps track of the rule activation correspondence:

$$l_{tr} \in L_{tr} \stackrel{\text{def}}{=} (El_{left}, El_{right}, El_{comp}) \quad (11)$$

*Note:*
The merge operation is generally considered as a primitive operation that combines contents of two source model elements (belonging to the left and right models) in order to produce one composed model element (without omission nor duplication of information) [7]. Therefore, the number of elements of each set defining a merging link $(El_{left}, El_{right}, El_{comp})$ is exactly 1.

$$l \in L_{merge} \Rightarrow |l.El_{left}| = 1 \wedge |l.El_{right}| = 1 \wedge |l.El_{comp}| = 1 \quad (12)$$

We want to weave relationships between traceability links (essentially to represent the rule calls sequence). Thus, each traceability relationship is defined as a couple connecting a parent link $l_{parent}$ with a child link $l_{child}$.

**Definition 12: Traceability relationship**
Let r be a traceability relationship. r is defined by:

$$r \in R_{tr} \stackrel{\text{def}}{=} (l_{parent}, l_{child}) \quad l_{parent}, l_{child} \in L_{tr} \quad (13)$$

*b) Traces generation*

The trace model $M_{trace}$ is generated as an additional target model of the composition operation. Thus, we have:

$$M_{trace} \in \{M_{comp_i} | i \in \{1,2\}\} \quad (14)$$

A traceability link is associated with each rule activation (in keeping with the type of the activated rule). This link keeps track of the activation effects by connecting the source elements captured by the rule application with the target elements which are produced. We note by $l_a$ the traceability link corresponding to the activation $a$, where:

$$\forall a \in Act_{spec}, \exists! \, l_a \in L_{tr} \quad (15)$$

**Definition 13: Traceability link corresponding to rule activation**
Let $a$ be an activation of a composition rule R. We define the corresponding traceability link by:

$$l_a \stackrel{\text{def}}{=} \left\{ l \; \middle| \; \begin{array}{l} l \in L_{tr} \wedge (a.El_{left} = l.El_{left}) \wedge \\ (a.El_{right} = l.El_{right}) \wedge (a.El_{comp} = l.El_{comp}) \\ \wedge ((a.R.type = \text{"merge"} \wedge l \in L_{merge}) | \\ (a.R.type = \text{"transform"} \wedge l \in L_{transf})) \end{array} \right\} \quad (16)$$

Each explicit call to a composition rule $Rc$ ($appelExp_{Rc}$) is traced by weaving a traceability relationship between the link corresponding to the current activation (the activation of the calling rule) and the one which refers to the triggered activation (the activation of the called rule). Thus the nesting of traceability links will be closely modeled on explicit rule calls sequence.

$$\left[ \begin{array}{c} \forall appelExp_{Rc}, \exists! \, r \in R_{tr}, \\ (r.l_{parent} = l_{appelExp_{Rc}.activ}) \wedge \\ (\exists a \in Act_{spec}, (a.R = Rc) \wedge (a.El_{letf} = appelExp_{Rc}.El_{left}) \wedge \\ (a.El_{right} = appelExp_{Rc}.El_{right}) \wedge (r.l_{child} = l_a) \end{array} \right] \quad (17)$$

Similarly, to each implicit call to a composition rule, we weave a traceability relationship between two links. The first captures the current activation, while the second one corresponds to the activation producing the target equivalents of the source elements to resolve ($resolvedEl$).

$$\forall appelImp, \exists! r \in R_{tr},$$
$$[(r.l_{parent} = l_{appelImp.activ}) \wedge$$
$$(\exists a \in Act_{spec},$$
$$(a.El_c = resolve(appelImp.resolvedEl))$$
$$\wedge (r.l_{child} = l_a))] \quad (18)$$

## 4. Traceability approach

In this section, we provide an overview of our traceability approach dedicated to the model composition operation. The way the corresponding traces are captured and structured is based on the formalization we proposed in the previous section. Indeed, the trace model is generated as an extra output model of the composition specification. For this purpose, the latter has to be modified to an equivalent version that includes the traces generation patterns. We consider this concern a cross-cutting concern, and we encapsulate it in a traceability aspect. The weaving operation is performed through an aspect oriented modeling approach [8] implemented with a set of graph transformations [9]. Traces thus generated will conform to a generic traceability metamodel which defines the traceability information formalized in Section 3.2.

### 4.1 Traceability metamodel

The definition we propose for a trace model includes two types of traceability information: links and relationships (Eq. 10). Drawing on this, the traceability metamodel depicted in Fig. 3 defines the *TraceLink* concept which represents a correspondence between a set of source elements (referenced by the *left* and *right* properties) and target elements (Eq. 11). Besides, it specifies the two types of traceability links that were introduced to express the composition behaviors in a trivial manner: merging links and transformation links.

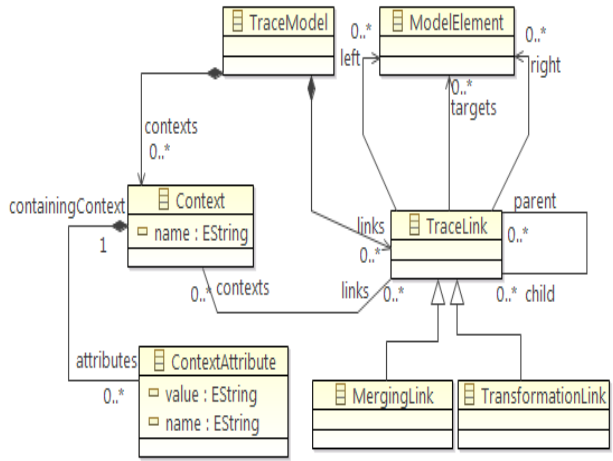

Fig. 3 The composition traceability metamodel.

As for traceability relationships (Eq. 13), they are expressed through parent-child relations among traceability links. The nesting of traces reveals the rule calls sequence (implicit and explicit calls), and it allows exploring traceability information at different granularity levels. Besides, the *Context* and *ContextAttribute* concepts provide the support to extend our traceability metamodel. We note that these concepts are not within the scope of the presented formalization. They allow the assignment of further expressiveness information to a subset of traceability links [2].

### 4.2 Generation mechanisms

We recall that the traces generation concern is encapsulated in a traceability aspect. However, the weaving operation is performed in the model level rather than directly manipulating the concrete composition specification. The objective is to define a generic traceability aspect around the formalization we proposed for a composition specification (cf. Section 3.1) while abstracting specificities of the language used to express it.

However, the Aspect-Oriented Modeling (AOM) is presented as a paradigm for encapsulating aspect, at the model level, which requires an implementation mechanism for the weaving operation. In the literature, two solutions have been presented: the composition of the base model with the aspect model [8]; and the simulation of aspect orientations by means of graph transformations [10]. In our approach, we opted for the second technique as it supports a better configuration of the aspect application.

Besides, in order to specify graph transformations which implement the weaving mechanism independently from a given composition language, these transformations have to be expressed around a pivot language. In [11], we proposed a generic composition language to deal with this issue. Basically, the metamodel definition is based on the formalization we presented for a composition specification (cf. Section 3.1), and it includes other concepts necessary to perform the weaving. Fig. 4 depicts an excerpt of this metamodel.

- *Source and target models*: The composition operation is defined as an operator that is applied on the source models in order to produce the composed ones (Eq. 1). The concept *Model* refers to the managed models, and describes also their metamodel.
- *Merging rules*: The *MergeRule* concept represents the first class of composition rules that were discussed in Section 3.1. It is connected to a set of parameters (*ParameterDec*) referencing the contributing elements, and has a statement block (StatementBlock) which specifies the merging mechanism.
- *Transformation rules*: As for transformation rules, they are defined by the *TransformationRule* concept.

Their structure is similar to merging rules, but no restriction is specified for the number of source and target parameters.
- *Rule call*: The expression of implicit rule calls exploits predefined operations that allow the resolution of target equivalents. We defined this operation as a mapping between source and target equivalents (Eq. 9). Within our generic composition language, this behavior is encapsulated in an abstract operation named *targetEquivalent*() [11].

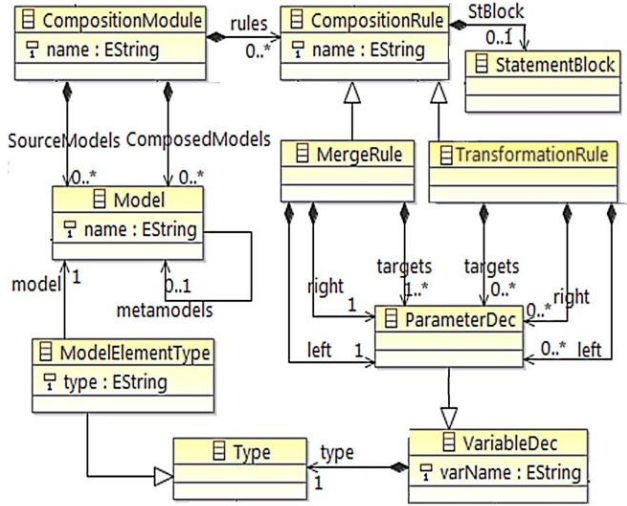

Fig. 4 Excerpt of the generic composition metamodel.

Based on this proposal, we have implemented a graph transformation unit for weaving the traces generation patterns (see Fig. 5). Its first rule declares the trace model to be generated as an additional target model of the composition to trace. Afterwards, the two following transformations are applied in order to trace all the merging and transformation rules contained in the specification.

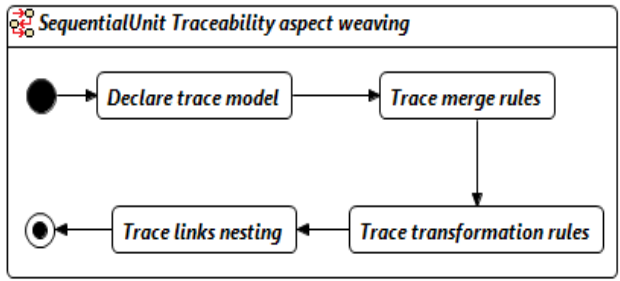

Fig. 5 Traceability weaving unit.

Basically, tracing a composition rule consists of adding a new parameter to its definition. This parameter references the traceability link to be generated each time the rule is activated. Besides, traceability information has to be assigned to it by initializing the *left*, *right*, and *targets* properties. Finally, the patterns responsible of nesting traceability links are woven by applying the last rule. The latter expresses the nesting mechanisms we presented in Section 3.2. More details about the implementation of these graph transformations are given in a previous work [11].

## 5. Related work

The execution of EML specifications relies on the exploitation of implicit traceability links. Indeed, all correspondences established in the comparison phase are stored in a trace model called *MatchTrace* [12]. This model is exported by the EML engine, and allows the latter to switch the application of transformation and merging rules on the appropriate elements. In adequacy with rule types, the corresponding traceability metamodel expresses two categories of traceability links. In addition, two mechanisms are provided to access these implicit traces [6]: The first is to export the trace model, and the second allows their exploitation through the *equivalent*() and *equivalents*() operations.

Similar to EML, the ATL language captures traces of rules applications (specifically declarative rules) in an implicit trace model. These traces provide a basis for the resolution of target elements produced by transforming certain source elements (through the *resolveTemp*() operation). On the other hand, *Jouault* [13] proposed an explicit traceability approach for ATL. An ATL module is considered as a model on which a higher order transformation is applied for embedding the traces generation code. The aim is to generate at each execution of a rule, a traceability link that connects the manipulated source elements with target ones. However, all correspondences between source and target model elements are captured as simple traceability links without any specialization.

The nesting of traces is another important aspect which has not been addressed by the aforementioned traceability solutions. This structure must be woven between traceability links to represent the rules invocation sequence. On the other side, the generic traceability framework proposed by *Grammel and Kastenholz* [14] can be specialized in order to trace ATL and EML specifications. Indeed, their traceability metamodel provides solutions for structuring traces. However, mechanisms for capturing such traceability information are not described. Also, concerning model composition scenarios, a lack of expressiveness of the generated traces can always be perceived.

# 6. Conclusion

This work aimed to formalize the model composition traceability. Indeed, we proposed a formal definition of the model composition operation, and we expressed, in a rigorous manner, the corresponding traces and the generation mechanisms. Furthermore, we provided an overview of our generic traceability approach dedicated to the composition of models. The design of this approach has taken up the presented formalization.

Indeed, our approach is based on the definition we proposed for the composition specification, and aims to capture two classes of traceability information. The first refers to the construction of the composed model by generating for each composition rule activations (leading to the production of a target element) a traceability link which captures its effect. The second type of information consists of traceability relationships. They are woven between traceability links to represent rule calls sequence. Such information reveals the way the composed model is structured. In addition, traceability links are categorized in merging and transformation links to resume the behaviors expressed by the traced rules.

Nevertheless, we recognize that the definition we proposed for a composition specification is not complete. For example, the rule body can express the construction of target model elements other than those referenced by its parameters. Likewise, the weaving of structural relationships between the composed model elements can be expressed directly without exploiting predefined operations to resolve target equivalents. Thus, our proposal should be extended to cover the missing elements, and the generation mechanisms should be adapted to capture the corresponding traces. The objective is to ensure a deeper analysis of the specification in order to generate more complete traces.